\documentclass[11pt]{iopart}
\usepackage{graphicx}
\begin{document}

\title{Stabilization of a laser on a large-detuned atomic-reference frequency by resonant interferometry}

\author{Priscila M. T. Barboza$^1$, Guilherme G. Nascimento$^1$, Michelle O. Ara\'ujo$^2$\footnote{Present address: Institut Non Lin\'eaire de Nice, CNRS and Universit\'e Nice Sophia-Antipolis, 1361 route des Lucioles, 06560 Valbonne, France.}, C\'icero M. da Silva$^4$, Hugo L. D. de S. Cavalcante$^3$, Marcos Ori\'a$^{2,4}$, Martine Chevrollier$^{2,4}$, Thierry Passerat de Silans$^{1}$}

\address{$^1$ Departamento de F\'isica-CCEN, Universidade Federal da Para\'iba - Caixa Postal 5008, 58051-970 Jo\~ao Pessoa, PB, Brazil.}
\address{$^2$ Laborat\'orio de Espectroscopia \'Otica, DF - CCEN, Universidade Federal da Para\'iba - Caixa Postal 5086, 58051-900 Jo\~ao Pessoa, PB, Brazil.}
\address{$^3$ Departamento de Inform\'atica, Centro de Inform\'atica, Universidade Federal da Para\'iba, Avenida dos Escoteiros -s/n$^\circ$ Mangabeira VII, 58055-000 Jo\~ao Pessoa, PB, Brazil.}
\address{$^4$ Universidade Federal Rural de Pernambuco, UACSA, BR 101 Sul, no 5225, Ponte dos Carvalhos, Cabo de Santo Agostinho, PE 54580-000, Brazil }
\ead{thierry@otica.ufpb.br}

\begin{abstract}
We report a simple technique for stabilization of a laser frequency at the wings of an atomic resonance. The reference signal used for stabilization issues from interference effects obtained in a low-quality cavity filled with a resonant atomic vapour. For a frequency detuned at 2.6 GHz from the $^{133}$Cs D$_2$ 6S$_{1/2}$ F=4 to 6P$_{3/2}$ F'= 5 transition, the fractional frequency Allan deviation is $10^{-8}$ for averaging times of 300 s, corresponding to a frequency deviation of 4 MHz. Adequate choice of the atomic density and of the cell thickness allows locking the laser at detunings larger than 10 GHz. Such a simple technique does not require magnetic fields or signal modulation.
\end{abstract}
\maketitle

\section{Introduction}
Atomic physics experiments frequently require the laser frequency to be stable for long periods of time. Many different techniques generate a reference signal to stabilize the laser frequency around the atomic resonance, within a range that is of the order of the Doppler width. Examples of such techniques are modulation of a linear absorption signal \cite{Yanagawa84}, DAVLL \cite{Corwin98} and ANGELLS \cite{Queiroga12,Martins12}.  However, in some experiments it is desirable to stabilize the laser frequency up to a few GHz from the line-center. Examples of such experiments are the search for generation of quantum-correlated twin beams in atomic vapours \cite{Guo14} and the control of atomic motion using light-gradient forces \cite{Ovchinnikov91}, where light absorption and subsequent spontaneous emission must be avoided. The usual way to stabilize the laser frequency in the wings of the resonance is to control the frequency difference between the main laser source used in the experiment and a second, auxiliary laser, locked to an atomic resonance. The frequency difference between the two lasers can be controlled using, for example, an external optical cavity \cite{Hansch80,Reich86} or a beating signal \cite{Schunemann99}. The experimental complexity of such techniques encourages the search for a reference signal originating from the wings of the atomic resonance itself, as the one presented here.

To the best of our knowledge, only two techniques allow one to obtain a reference signal directly from the atomic vapour, without the need of a second oscillator. Marchant and co-authors \cite{Marchant11} used a high magnetic field to induce a Faraday rotation in a laser beam that is transmitted through an atomic vapour. The imbalance of the transmitted power in two orthogonal polarization components is used as the reference signal. Such transmission imbalance originates in the polarization rotation with an angle that is proportional to the derivative of the refractive index \cite{Kemp11}. The reference signal is shown to exhibit a spectral oscillatory behaviour in the wings of the resonance and is used to stabilize the laser frequency. Sargsyan and co-authors \cite{Sargsyan14} used a strong magnetic field ($B>20$ mT) in a micrometer-thin cell to displace the atomic resonance by a few GHz and performed a saturated absorption on this displaced resonance to stabilize the laser frequency. Here we describe a magnetic field-free technique to obtain a reference signal directly from a single laser beam transmitted through a warm atomic vapour. The reference signal is generated by the interferometric transmission of the parallel windows of the optical cell, modulated by the frequency-dependent refractive index of the resonant vapour inside the cell. We use the resulting ``oscillatory'' behaviour  at the resonance wings to stabilize the laser against frequency drifts.

\section{Experimental set-up}
\begin{figure}[hbtp]
\centering 
\fbox{\includegraphics[width=0.85\linewidth]{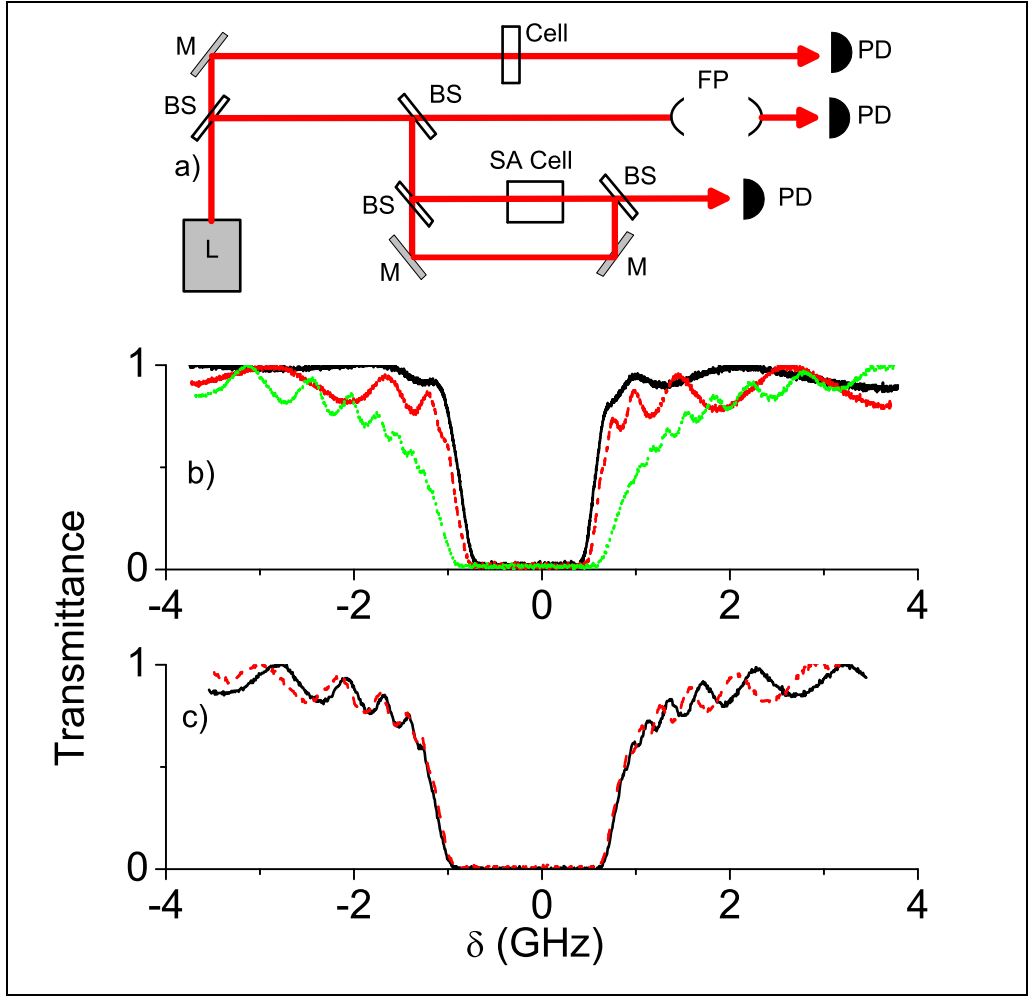}}
\caption{(Color online)(a) Experimental scheme: a laser beam is sent through a 1-mm-thick vapour cell. Auxiliary saturated-absorption and Fabry-P\'erot are set-up to calibrate the laser frequency. L is the diode laser; M are mirrors; BS are beam-splitters; FP is a Fabry-P\'erot interferometer; SA Cell is the cell where saturated-absorption is performed; PD are photodetectors and Cell is the cell used to obtain the reference signal. (b) Experimental transmittance as a function of frequency detuning for different atomic densities: $6\times10^{13}$ atoms/cm$^{3}$ (black - solid line), $1.4\times10^{14}$ atoms/cm$^{3}$ (red - dashed lined) and $3.5\times10^{14}$ atoms/cm$^{3}$ (green - dotted line). (c) Experimental transmittance as a function of frequency detuning for different incidence angles: normal incidence (black - solid lines) and 40 mrad (red - dashed line). Zero detuning corresponds to the $6S_{1/2}(F=4)\rightarrow 6P_{3/2}(F'=5)$ hyperfine transition.}
\label{Fcell}
\end{figure}

The experimental set-up consists in detecting the transmission of a low-power laser beam at near normal incidence through an optical cell containing an atomic vapour (Fig. \ref{Fcell}a), in our case a caesium vapour. When the atomic density is high enough (typically with total absorption at line center) the transmittance as a function of the frequency exhibits oscillations in the wings of the resonance (Fig. \ref{Fcell}(b,c)). The slopes of this oscillatory signal can be used to generate an error signal to stabilize the laser frequency.\\

The laser used in our experiment is a distributed feedback (DFB) semiconductor emitting at 852 nm, resonant with the D2 line of caesium. A small fraction of the laser power is sent to an auxiliary set-up for measurements of saturated absorption and transmission of a Fabry-P\'erot (FP) cavity, both to monitor the laser frequency. We use the saturated absorption spectra to determine the zero of the detuning axis (arbitrarily chosen to be the 6S$_{1/2}(F=4)\rightarrow 6P_{3/2}(F'=5)$ transition) and the 1.5 GHz-free-spectral-range FP interferometer to determine the detuning of the locking point in relation to this zero. A low-power, collimated and linearly polarized laser beam (20 $\mu$W - diameter of 5 mm) is sent to the stabilization cell. The low intensity of the beam ( $I=0.1$ mW/cm$^2$ is much less than the saturation intensity, $I_s\approx1.6$ mW/cm$^2$) assures a linear regime for the interaction with the vapour, even at the center of the resonance.

The cell is T-shaped with a thickness of 1mm and a window parallelism within 5 mrad. The atomic density is controlled by the temperature of a side-arm Cs reservoir, which contains a Cs drop. The cell body is also heated, but with a temperature higher than the reservoir to avoid Cs condensation on the windows. Both the cell-body oven and the reservoir oven are heated by copper wires making clockwise and counter-clockwise loops. Such configuration of the copper wire avoids the generation of significant magnetic fields. We have checked the influence of a possibly residual magnetic field by turning off the current in the heater wires and observing that the transmittance does not change.

\section{Origin of the reference signal}
The cell with almost parallel windows acts as a resonator of poor finesse. For an empty cell with thickness of $1$ mm, the cavity free-spectral-range is of 150 GHz. This range is much larger than the oscillatory behaviour observed in Fig. \ref{Fcell}(b,c), because the vapour inside the cavity has a refractive index $n_V$ that changes with the laser detuning ($\delta$) from resonance, $n_V(\delta)=1+\delta n_V(\delta)$. The phase acquired by the field in one passage through the cell $\phi=\frac{2\pi}{c} n_V\frac{L}{\cos(\theta)}\left(\nu_0+\delta\right)$ might rotate by $2\pi$ even for small frequency changes. In the above expression, $\theta$ is the angle of incidence of the beam on the cell windows, $L$ is the cell thickness, $c$ is the speed of light in vacuum and $\nu_0$ is the frequency of the atomic transition.

To calculate the transmitted beam in the cell cavity, we have developed a model taking into account the atomic vapour refractive index and its absorption. Our calculations consist in summing the fields resulting from multiple reflection. The transmittance takes into account the field attenuation $e^{-\frac{1}{2}\alpha L}$, where $\alpha$ is the vapour absorption coefficient, and the phase $e^{i\phi}$, that is acquired at each passage. The resulting transmittance is:

\begin{equation}
T=\frac{\left(1-R^2\right)e^{-\alpha L}}{\left(1-Re^{-\alpha L}\right)^2+4Re^{-\alpha L}sin^2\left(\phi\right)}, \label{theorietrans}
\end{equation}

with $R$ the reflectance of the windows. For quartz windows at normal incidence, we use $R\approx 3.5$ \%.

The transmittance in Eq. \ref{theorietrans} exhibits maxima when the phase is $j\pi$, with $j$ an integer. As $dn_V/d\delta\propto N$, the separation between adjacent maxima decreases with increasing density. This is, indeed, observed experimentally as shown in Fig. \ref{Fcell}b where the transmittance for different atomic densities is plotted. Also, the separation between maxima and the frequency at which they occur should depend on the incident angle, as experimentally observed in Fig \ref{Fcell}c.

\section{Laser frequency stabilizationn}
Once chosen a given configuration of incidence angle, incidence position and vapour density, the oscillatory transmittance of the laser beam through the cell can be used to generate an error signal to stabilize the laser frequency. The error signal is generated by adding a controllable voltage offset to the transmission signal in order to achieve a zero value at the frequency desired for stabilization. The error signal is integrated and is sent to the current control of the laser to correct frequency drifts. 

\begin{figure}[hbtp]
\centering
\fbox{\includegraphics[width=\linewidth]{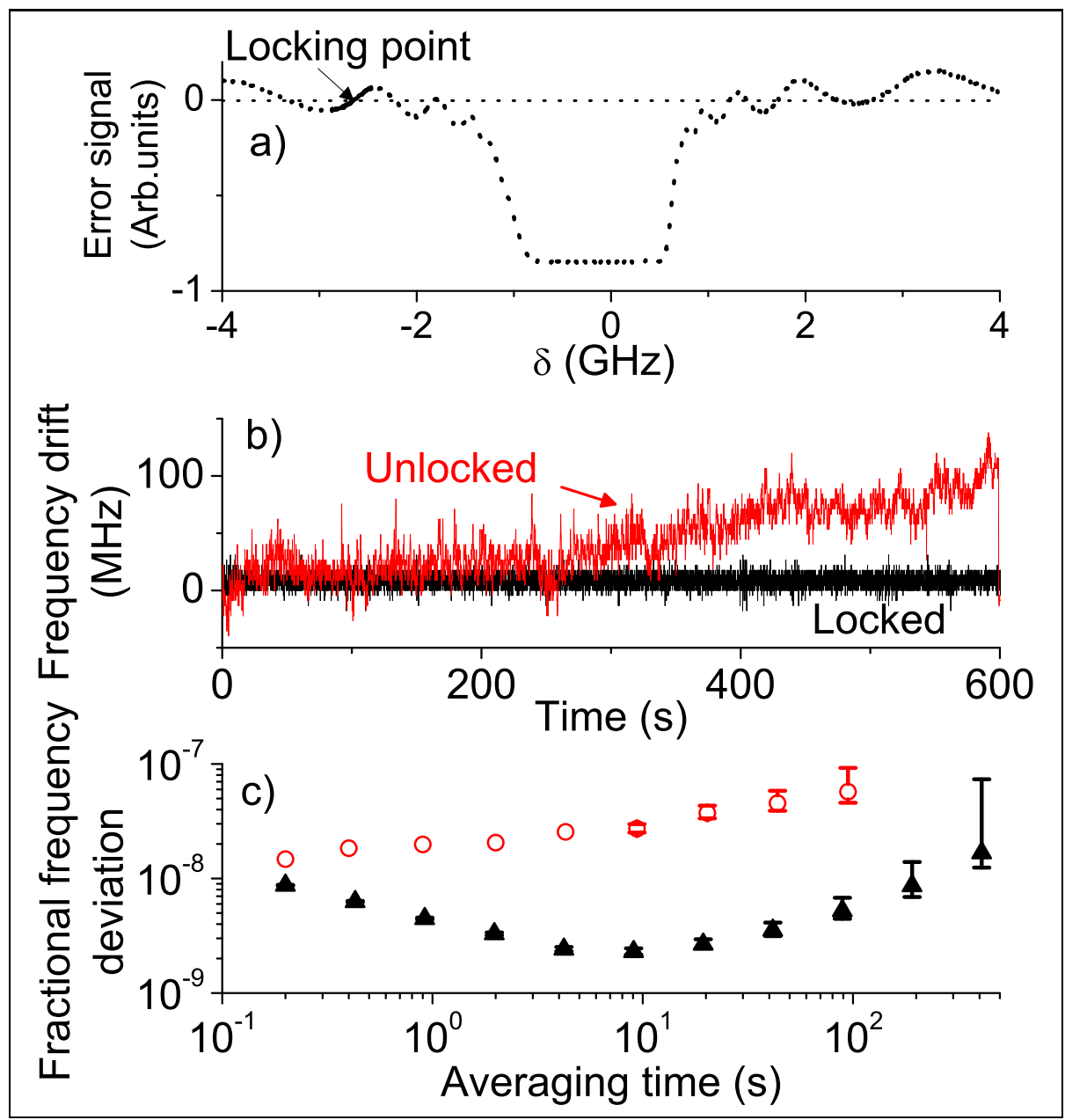}}
\caption{(Color online) (a) Error signal for a particular locking point at around $\delta=-2650$ MHz from line center. The solid line evidences the capture range for this locking point. This particular error signal was obtained for an atomic density of $N=2\times10^{14}$ atoms/cm$^3$ and the laser beam at normal incidence. (b) Frequency drift relative to the locking point at $\delta=-2650$ MHz for a locked and for an unlocked laser during 10 minutes. (c) Allan deviation calculated from time series of Fig. \ref{TR6}b for locked (black triangles) and unlocked laser (red circles).}
\label{TR6}
\end{figure}

Any slope of the oscillating transmittance can be used to stabilize the laser frequency. For instance, with an atomic density of $N=2\times10^{14}$ atoms/cm$^3$, it is possible to stabilize the laser from $\delta\sim1000$ MHz to at least, $\delta=3000$ MHz (see Fig \ref{TR6}a). The laser frequency drift for a frequency stabilized at detuning -2650 MHz is shown in Fig. \ref{TR6}b together with the non-stabilized case. The laser frequency drift, shown in figure \ref{TR6}(b,c), was monitored from the error signal itself for a locked and for an unlocked laser. Although such a method of monitoring the laser drift is not an absolute one, it testifies to the usefulness of the proposed scheme to stabilize the laser frequency. The stabilization feedback eliminates the laser drift, that changes no more than 20 MHz peak-to-peak (7 MHz of rms fluctuation) in 10 minutes. We have observed locking times longer than one hour (no long run data acquired) and similar performance is obtained for all reference frequencies corresponding to the several slopes of Fig. \ref{TR6}a.

From the time series of Fig. \ref{TR6}b, obtained from the error signal, we calculate the Allan variance using \cite{Hall}:
\begin{equation}
\sigma^2_y(\tau)=\frac{1}{2}\frac{1}{M-1}\sum_{i=1}^{M-1}\left[\bar{y}_{i+1}-\bar{y}_i\right]^2,
\end{equation}
where $M$ is the number of time series of duration $\tau$ in which the original time series can be divided and $\bar{y}_i$ is the average of the fractional frequency over a time $\tau$ for the i-th time series. The Allan deviation ($\sigma_y(\tau)=\sqrt{\sigma^2_y(\tau)}$) is plotted in Fig. \ref{TR6}c as a function of the averaging time. For averaging times up to 10 s, $\sigma_y(\tau)$ decays as $\tau^{-1/2}$ as expected for white noise. For an averaging time of 5 s, $\sigma_y=2\times10^{-9}$ (corresponding to a noise width of 1 MHz), while for an averaging time of 300 s we have $\sigma_y=10^{-8}$, corresponding to a 4 MHz deviation from the mean locking point. Notice that the performance obtained with the technique proposed here is very similar to that reported in \cite{Marchant11}. The rather large fractional frequency for small time averaging results from the fact that the laser we have used has a rather broad linewidth and that our home-made electronics are not fast enough to correct for these short-time fluctuations ($<1$ s).

In order to stabilize the laser frequency for long-running experiments, it is important that the reference signal remains stable over time. Reference signal fluctuations are mainly attributed to the variations of the atomic density through changes of the reservoir temperature. Fluctuations of the atomic density proportionally affect the vapour refractive index and, thus, the phase accumulated by the field passage through the cell.  From our theoretical model, a variation of 1 $^\circ$C of the reservoir temperature, corresponding to a change of 5\% of the atomic density in the range of densities we work with, results in a displacement of 150 MHz of a locking point at -3000 MHz (locking point displacement of $0.05\delta/^{\circ}$C). Notice that this is a frequency stability similar to that obtained in \cite{Marchant11} where a variation of 200 MHz/$^\circ$C was obtained for a locking point at 10 GHz from line center. For the home-made oven we use, the locking point at $\sim-2650$ MHz drifted by approximately 100 MHz in one hour, corresponding to an estimated change of the reservoir temperature of 0.8 $^\circ$C. An active control of the oven heater current, and thus of the cell reservoir temperature, should allow a much more stable reference signal. For instance, the stabilization of the reservoir temperature to within 1mK should allow the locking point to displace by only fractions of MHz. Change in the cell thickness due to thermal expansion is negligible. Indeed, for the fused-silica cell that we use, the linear expansion coefficient is $\gamma\sim 10^{-6}$ K$^{-1}$ and cell dilatation results in a displacement of the slope of the error signal of $\sim 3$ MHz/$^\circ$C. As evidenced in Fig. \ref{Fcell}c, another source of the reference signal instability is a mechanical change of the incidence angle. The main effect of the incident angle in the signal is to change the effective cell thickness $L\rightarrow L/cos(\theta)$. For a slope around $\delta=-3000$ MHz, a change of the reference of $\sim120$ MHz would be expected if the incident angle were modified 10 mrad from the normal incidence.

\begin{figure}[hbtp]
\centering
\fbox{\includegraphics[width=\linewidth]{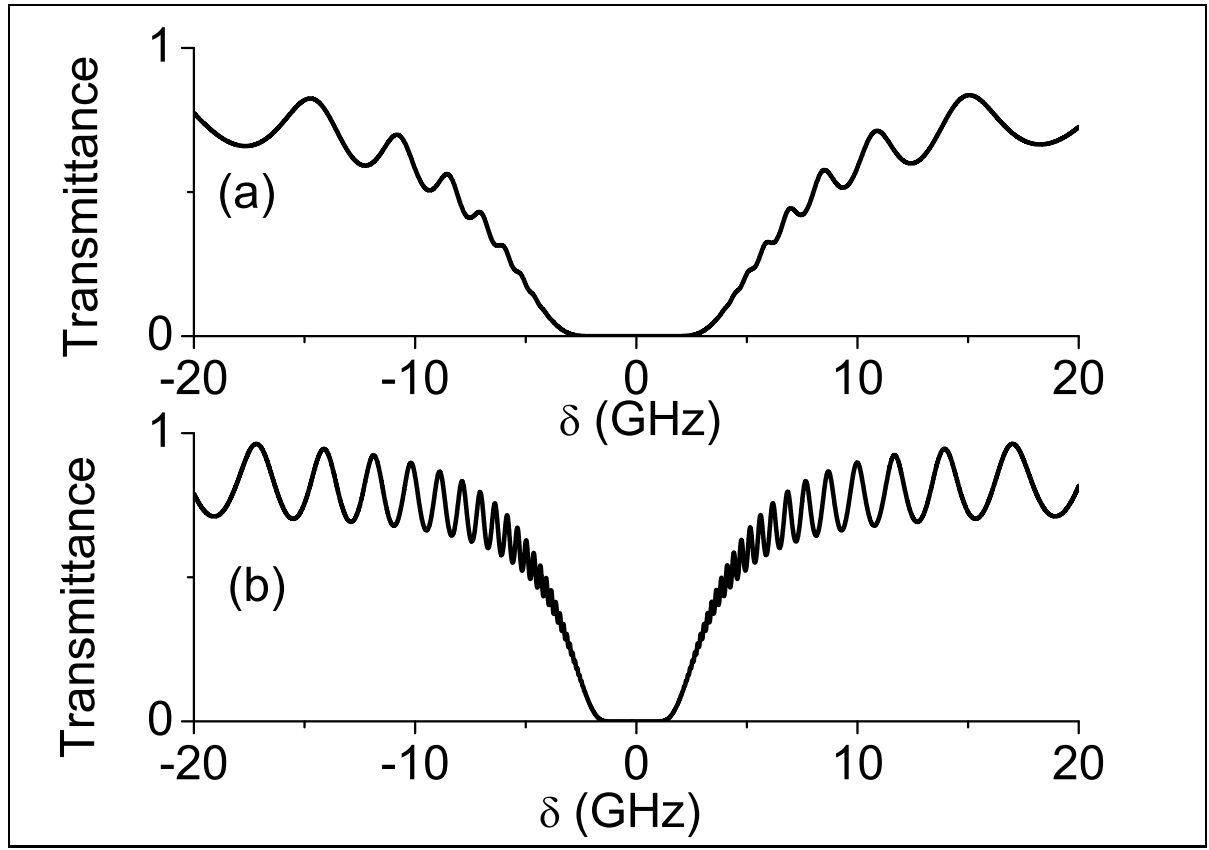}}
\caption{Transmittance calculated from Eq. \ref{theorietrans} for: (a) a 1-mm thickness cell and a atomic density of $N=8.5\times10^{14}$ atoms/cm$^3$; (b) a 10-mm thickness cell and an atomic density of $N=1.3\times10^{14}$ atoms/cm$^3$.}
\label{errorT}
\end{figure}

\begin{figure}[hbtp]
\centering
\fbox{\includegraphics[width=\linewidth]{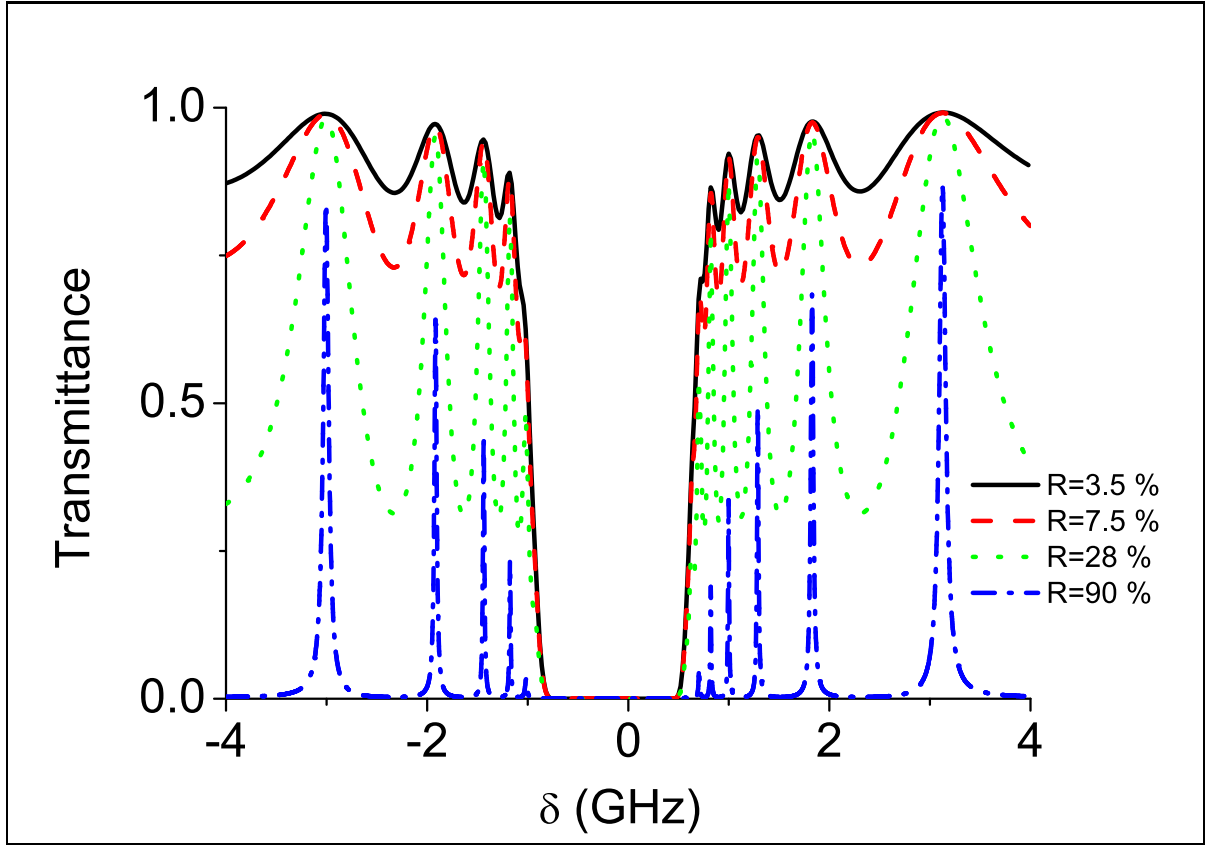}}
\caption{(Color online) Transmittance calculated from Eq. \ref{theorietrans} for different windows reflectances, for a 1-mm thick cell and an atomic density of $N=1.3\times10^{14}$ atoms/cm$^3$.}
\label{Qfactor}
\end{figure}

For the atomic density that we used to obtain Fig. \ref{TR6}, locking points in both sides of the resonance in the range $1000$ MHz$\leq\left|\delta\right|\leq3000$ MHz are available. Although the stabilization points lie on slopes of the oscillatory signal, the position of those slopes can be slightly adjusted by changing either the temperature of the reservoir or the incident angle, allowing to lock the laser frequency anywhere in this 2 GHz range. Moreover, the range where stabilization points can be obtained can be increased by taking higher atomic densities or by using thicker cells, as illustrated in Fig. \ref{errorT}. Indeed, we have calculated (Eq. \ref{theorietrans}) the transmittance signal for a 1mm-thick cell and an atomic density of $N=8.5\times10^{14}$ atoms/cm$^3$ and we observe slopes of the oscillating signal for detunings $\delta>10$ GHz (Fig. \ref{errorT}a). The strategy of using higher atomic densities, i.e., higher reservoir temperatures, must be taken with care since alkali metals can react with fused silica at those temperatures. Cells bodies made of YAG or sapphire would be more appropriate in this case. Alternatively, locking points for $\delta>10$ GHz can also be obtained for an atomic density $N=1.3\times10^{14}$ atoms/cm$^3$ and a cell thickness of 1 cm (Fig. \ref{errorT}b). The strategy of using longer cells renders steeper slopes, which may improve the locking performance, at the expense of the stability of the reference signal since $d\phi\propto dnL\propto \ dNL$. In our experimental set-up we used a cell with quartz windows with reflectance $R=3.5$ \%. Increasing the windows reflectance, and thus the cavity Q-factor, decreases the width of the cavity resonances and consequently increases the slope of the reference signals. We have calculated the reference signal for different values of $R$ and show the results in Fig. \ref{Qfactor}. The position of the maxima does not depend on the windows reflectance since they are governed by $\phi=j\pi$. As expected, the slopes increase as $R$ increases, improving the locking performance, but at the expense of simplicity. Notice that the use of a cell with sapphire windows ($n_S=1.76$, $R=7.5$ \%) results in a gain of a factor 2 in the signal slopes and allows one to work with higher temperatures. In the calculations of Fig. \ref{Qfactor} we have considered linear interactions between the laser beam and the vapour to determine $\alpha$ and $n_V(\delta)$. However, for high values of $R$, field enhancement inside the cavity might turn the interaction into nonlinear, resulting in a displacement of the maxima.

\section{Summary}
We have developed a new technique that allows the stabilization of the frequency of a laser in the wings of an atomic resonance. The technique explores a frequency-dependent signal exhibiting spectral oscillations that originate from multiple reflections between the windows of the cell combined with a frequency-dependent refractive index due to the atomic vapour inside the cavity. Specifically, using a 1mm-thick cell and atomic densities around $2\times10^{14}$ atoms/cm$^{3}$, we were able to lock the laser frequency in a range from 1.0 GHz to 3.0 GHz from line center, with a fractional frequency Allan deviation of $\sigma_y$=2$ \times 10^{-9}$ (fluctuations of 1 MHz) for an integration time of 5s and $\sigma_y= 10^{-8}$ (deviations of 4 MHz) for an integration time of 300 s. The main source of instability of the reference signal is fluctuation of the atomic density through modifications of the temperature of the Cs reservoir. We estimate a displacement rate due to fluctuations of the reservoir temperature of $0.05\delta$/$^\circ$C, similar to the technique reported in \cite{Marchant11}. The higher Fractional Frequency deviation that we have obtained relative to \cite{Marchant11} is attributed to two factors: i) to the fact that the electronic feedback circuit is slow (an integrator with time constant of tens of seconds scales) and ii) to drift of the locking point due to fluctuation of the reservoir temperature of the order of 3 Mhz in a time scale of 100s. Our theoretical model indicates that the range of stabilization can be extended by using, for example, either higher atomic densities or thicker cells. Moreover, by using windows with higher reflectance, steeper slopes of the reference signal are obtained, which represents a possible route to improving the locking performance. We stress that to implement the technique described here, one does not need to know exactly how the refractive index of the vapour changes with frequency, but only guarantee that the field phase rotate by $\pi$ for one passage through the cell for moderate frequency sweeps. Changes in the refractive index possibly caused by collisions, by the presence of dimers or by impurities will only displace the reference signal, and a particular frequency for the reference slope may be reached by small adjustments of the reservoir temperature or of the incident angle. Notice that the simple experimental set-up of the technique reported here does not require magnetic field or modulation operations to generate a reference signal for laser frequency stabilization.

\ack
We thanks funding from Brazilian agencies \textit{Conselho Nacional de Desenvolvimento Cient\'ifico e Tecnol\'ogico} (CNPq), \textit{Financiadora de Estudos e Projetos} (FINEP) and \textit{Coordena\c c\~ao de Aperfei\c coamento de Pessoal de N\'ivel Superior} (CAPES).

\end{document}